\documentclass[letterpaper]{article}
\usepackage{aaai}
\usepackage{times}
\usepackage{helvet}
\usepackage{courier}

\usepackage{graphicx}
\usepackage{url}
\usepackage{subfig}
\usepackage{mathtools}
\usepackage{tabularx, multirow}
\usepackage{dblfloatfix}


\begin{document}

\title{Analysing Timelines of National Histories across Wikipedia Editions:\\ A Comparative Computational Approach}
 
\author{
Anna Samoilenko\textsuperscript{1,2}, Florian Lemmerich\textsuperscript{1,2},
Katrin Weller\textsuperscript{1},
Maria Zens\textsuperscript{1},
Markus Strohmaier\textsuperscript{1,2}   \\
      \textsuperscript{1} GESIS -- Leibniz-Institute for the Social Sciences\\
      \textsuperscript{2} University of Koblenz-Landau\\
}

\maketitle
\begin{abstract}
Portrayals of history are never complete, and each description inherently exhibits a specific viewpoint and emphasis. In this paper, we aim to automatically identify such differences by computing timelines and detecting temporal focal points of written history across languages on Wikipedia. In particular, we study articles related to the history of all UN member states and compare them in 30 language editions. We develop a computational approach that allows to identify focal points quantitatively, and find that Wikipedia narratives about national histories (i) are skewed towards more recent events (\textit{recency bias}) and (ii) are distributed unevenly across the continents with significant focus on the history of European countries (\textit{Eurocentric bias}). We also establish that national historical timelines vary across language editions, although average interlingual consensus is rather high.
We hope that this paper provides a starting point for a broader computational analysis of written history on Wikipedia and elsewhere.

\end{abstract}
%\keywords{Wikipedia | multilingual | quantitative historiography | Null Model | computational history | collective memory} 

%%%%%%%%%%%%%%%%%%%%%%%%%%%%%%%%%%%  Intro  %%%%%%%%%%%%%%%%%%%%%%%%%%%%%%%%%%%%%%%%%%
%%%%%%%%%%%%%%%%%%%%%%%%%%%%%%%%%%%%%%%%%%%%%%%%%%%%%%%%%%%%%%%%%%%%%%%%%%%%%%%%%%%%%%
\section{Introduction}
Writing about history -- historiography -- is important in all social groups. Establishing some consensus on relevant dates provides a feeling of roots, and is at the core of building identities -- for individuals, groups, or nations. Each description inherently presents a unique viewpoint on past events, and it might be partial and disputable.
Today, the online encyclopedia Wikipedia has a vast readership across continents and languages. It offers quick, effortless access to a spectrum of reference information, including historical accounts. These representations might contain errors and false information \cite{potthast2008automatic}, be biased towards specific viewpoints, or differ otherwise from the books written by professional historians.
Fortunately, Wikipedia's open and digital nature allows for thorough quantitative analysis of historical narratives, even across a large number of languages -- something which is not a typical case for other historiographical sources, such as printed encyclopedias or history textbooks.

In this paper, we investigate descriptions of national histories in different Wikipedia language editions, taking a comparative computational approach. In that direction, we pursue two goals: (1) presenting a data-driven approach that enables analysis of historiography through a computational lens, and (2) answering specific research questions on the depiction of history in Wikipedia.

\textbf{Approach.} We present a computational approach to the analysis of textual historiographical data which is suited for large-scale comparative studies. We apply it to Wikipedia articles on all UN member states in 30 language editions. We concentrate on \textit{year dates} as accessible representations of more complex historical structures. To be able to compare the descriptions across languages, we retrieve from article texts all date mentions (in the form of 4-digit numbers between 1000-2016), and use them as a language-independent unit of comparison \cite{rusen1996}.
We propose a simple randomisation technique to extract \textit{significant focal points of national histories} -- time periods of significantly high mentions, compared to a random expectation model. We combine visual interpolation and expertise of history experts in order to evaluate how the results of our approach compare with the existing historical knowledge. We use hierarchical clustering to group countries whose histories are represented similarly on Wikipedia. Finally, we compute inter-language agreement on history of each country using the Jensen-Shannon divergence measure.  

\textbf{Empirical questions.} 
We compare, what readers of different languages can learn about national histories from their `home' Wikipedia language editions.
In particular, we focus on three research questions:
\textbf{RQ1:} What are the most documented periods of history of the last 1,000 years in Wikipedia? \textbf{RQ2:} What are the temporal focal points in descriptions of national histories in Wikipedia? \textbf{RQ3:} Are country timelines consistent across language editions?

\textbf{Empirical findings.} 
We find the presence of recency bias across all language editions and countries -- the tendency to document recent events more frequently than those that happened in a more distant past. We also find that the distribution of historical focal points in the analysed articles is inhomogeneous across continents. 
We discover a multitude of focal points distributed through entire timelines of European countries,
while we see much fewer highlights in pre-Columbian Americas and Oceania. 
Groups of countries with similarly distributed focal points map well to geopolitical blocs. 
Finally, we find differences in the way national histories are described in the examined language editions, although on average the cross-lingual consensus is rather high. 

\textbf{Contributions.} We contribute to the pool of computational methods that help to quantify historiographical processes. We combine multiple computational methods into an approach that can be used for quantitative analysis of large textual historical and historiographical datasets, such as demographic and economic records, census data, digitised books, etc. Our approach scales well, and is suited for large comparative studies of multidimensional (e.g. multiple languages and countries) data. 

By including 193 countries and 30 languages, we step beyond the current state of comparative historiography and allow for a large-scale transnational perspective on similarities, conjunctions, or alternatives in historiography. Although we start from the (limited) concept of the (pre-)\-histories of nation-states we, (i) methodologically, enable cross-lingual and -national clustering and comparison and, (ii) empirically, show that historiographical focal points transcend national borders, and contribute to existing literature on collective memory and public history as created and perceived through Wikipedia.

%%%%%%%%%%%%%%%%%%%%%%%%%%%  Literature   %%%%%%%%%%%%%%%%%%%%%%%%%%%%%%%%%%%%%%%%%%%%
%%%%%%%%%%%%%%%%%%%%%%%%%%%%%%%%%%%%%%%%%%%%%%%%%%%%%%%%%%%%%%%%%%%%%%%%%%%%%%%%%%%%%%
\section{Related literature} 
Our approach carries characteristics of `the digital turn' that the study of history has envisaged in the recent years:
it uses a large (digital) data set, borrows from statistical methods, and conceptually, turns to transnational and global comparative perspectives. 
Theoretically, our approach lies in the domain of cultural history (analysing the multitude of historical interpretations, e.g. gender-based or post-colonial histories), with a specific focus on the formation and effects of collective/public memories \cite{conrad}, and the analysis of nations as imagined communities~\cite{anderson2016}.

\begin{table*}[!b]
\centering
\small
\caption{Expected error rates of dates extraction: language editions and centuries}
\label{table:ev}
\begin{tabular}{|lc|lc|lc|}
\hline
\multirow{2}{*}{Language} & \multirow{2}{*}{Exp. error.} & \multirow{2}{*}{Language} & \multirow{2}{*}{Exp. error.} & \multirow{2}{*}{Language} & \multirow{2}{*}{Exp. error.} \\
&&&&&\\
\hline
English    & 0.0067 & Ukrainian        & 0.0351 & Basque            & 0.0042 \\
German     & 0.0186 & Catalan          & 0.0096 & Bulgarian         & 0.0262 \\
Swedish    & 0.0109 & Norwegian Bokmal & 0.0538 & Danish            & 0.0086 \\
Dutch      & 0.0198 & Serbo-Croatian   & 0.0068 & Slovak            & 0.0040 \\
French     & 0.0018 & Finnish          & 0.0208 & Lithuanian        & 0.0266 \\
Russian    & 0.0395 & Hungarian        & 0.0347 & Croatian          & 0.0178 \\
Italian    & 0.0081 & Romanian         & 0.0378 & Slovenian         & 0.0025 \\
Spanish    & 0.0069 & Czech            & 0.0076 & Estonian          & 0.0131 \\
Polish     & 0.0223 & Serbian          & 0.0119 & Galician          & 0.0154 \\
Portuguese & 0.0166 & Turkish          & 0.0256 & Norwegian Nynorsk & 0.0246 \\
\hline
\end{tabular}
\begin{tabular}{|lc|}
\hline
Century & Exp. error. \\
\hline
11th century & 0.2428 \\
12th century & 0.0442 \\
13th century & 0.0982 \\
14th century & 0.0214 \\
15th century & 0.0363 \\
16th century & 0.0415 \\
17th century & 0.0261 \\
18th century & 0.0089 \\
19th century & 0.0094 \\
20th century & 0.0000 \\
21st century & 0.0100 \\
\hline
\end{tabular}
\end{table*}

\textbf{Wikipedia as a data source:} 
Many non-academics start seeing history as a venue for active participation, rather than a domain of professional historians~\cite{rosenzweig1998}.
The encyclopedia Wikipedia is open for everyone to contribute on any topic. Thanks to this feature, it has become one of the primary venues where `free-lancer' amateur historians, potentially, alongside with professionals, can participate in history-making and shaping historiographic discourse \cite{conrad}.
A number of professional historians recognise Wikipedia as a place where enthusiasts collaboratively re-think the past \cite{pfister}, construct public memories \cite{Pentzold-12}, and write history in an open source manner \cite{rosenzweig2006}. Although such popular understandings of the past might differ from those of professional historians \cite{conrad}, Wikipedia is a popular source of information when it comes to history \cite{spoerri2007} and thus has become an object of research itself.

To the best of our knowledge, only a few researchers have investigated historical narratives of Wikipedians: Luyt \shortcite{luyt} compared the articles on the history of two countries, concluding that the history of Singapore recounts the dominant political narrative, while the article on the history of the Philippines contains both traditional and alternative views. Jensen \shortcite{jensen2012} looked into the discussion pages about the article on the war of 1812 and found that the main debate among the editors was on who won the war. Both studies use a traditional descriptive methodology. Finally, Gieck et al. \shortcite{gieck} used a data science approach and compared war-related articles across five language editions, using methods from sentiment-, network-, and language complexity analysis. The authors found that World Wars I and II are the most important historical events in these editions.

\textbf{Quantifying history:}
Quantitative approaches were integrated into history studies in the last century. 
Opposite to traditional qualitative interpretations, they relied on statistical methods and a new conceptualisation, in which historical reality was condensed to quantifiable (often socioeconomic) historical facts, whose evolution was traced through longitudinal studies \cite{furet1971}.
Computational approaches that allow formulating and testing retrospective hypotheses, running historical experiments, and discovering large-scale patterns of the past by processing big data-sets, appear to be the obvious next step that could turn history into an analytical, deductive, predictive science \cite{turchin2011,kiser}. 
Mathematical simulations have helped to test historical hypotheses about the evolution of commodity flows across ancient Asia \cite{malkov}, the influence of agriculture on birth rates in the Old World \cite{bennett}, and the rise and fall of large-scale societies \cite{turchin2013}. Network approaches have also found popularity among historiographers, due to their possibility to visualise and quantify relational ties that are abundant in written (digitised) historical texts. Networks have been used to map the travels and settlement of the Vikings \cite{sindbaek}, to study centralisation of political parties in Renaissance Florence \cite{padgett1993}, to examine interconnections between the elite individuals of Medieval Scotland \cite{jackson}, and to track the intellectual mobility of notable individuals on a large scale \cite{schich}. 

The step from statistic to historic interpretation still remains a difficult one, which is one of the reasons quantitative approaches have been slow in gaining support among the traditional historians. Nevertheless, computational studies could contribute evidence to support the existing historical theories, or suggest otherwise unavailable new hypotheses. The recent rise of interest to digital humanities and successes in digitising large collections of historical documents \cite{Michel,yu2016pantheon} allow broad possibilities for historians to select the data relevant to their questions. 
Still, the pool of available methods remains rather sparse. As it becomes easier for historians to extract data from digitised records, as well as from digitally born sources such as Wikipedia, new methods are in need that will help quantify and map historical processes. 

%%%%%%%%%%%%%%%%%%%%%%%%%%%%%%%%%%%  Data  %%%%%%%%%%%%%%%%%%%%%%%%%%%%%%%%%%%%%%%%%%%
%%%%%%%%%%%%%%%%%%%%%%%%%%%%%%%%%%%%%%%%%%%%%%%%%%%%%%%%%%%%%%%%%%%%%%%%%%%%%%%%%%%%%%
\section{Data collection \& validation}
In this section we describe the steps of data collection and validation. Both data and code are available online \cite{dataset}.
%\footnote{The URL is hidden to preserve the anonymity of the authors.~\url{}} to ensure reproducibility. 
We focus on the history of 193 countries\footnote{Throughout the paper we use the terms nation, country, and state as synonyms, being aware of the differences.} which are the current UN member states\footnote{List of the UN member states,~\url{http://www.un.org/en/member-states/index.html} (accessed Nov.~13,~2016)}. 

\begin{figure}[t]
    \centerline{\includegraphics[width=.5\textwidth]{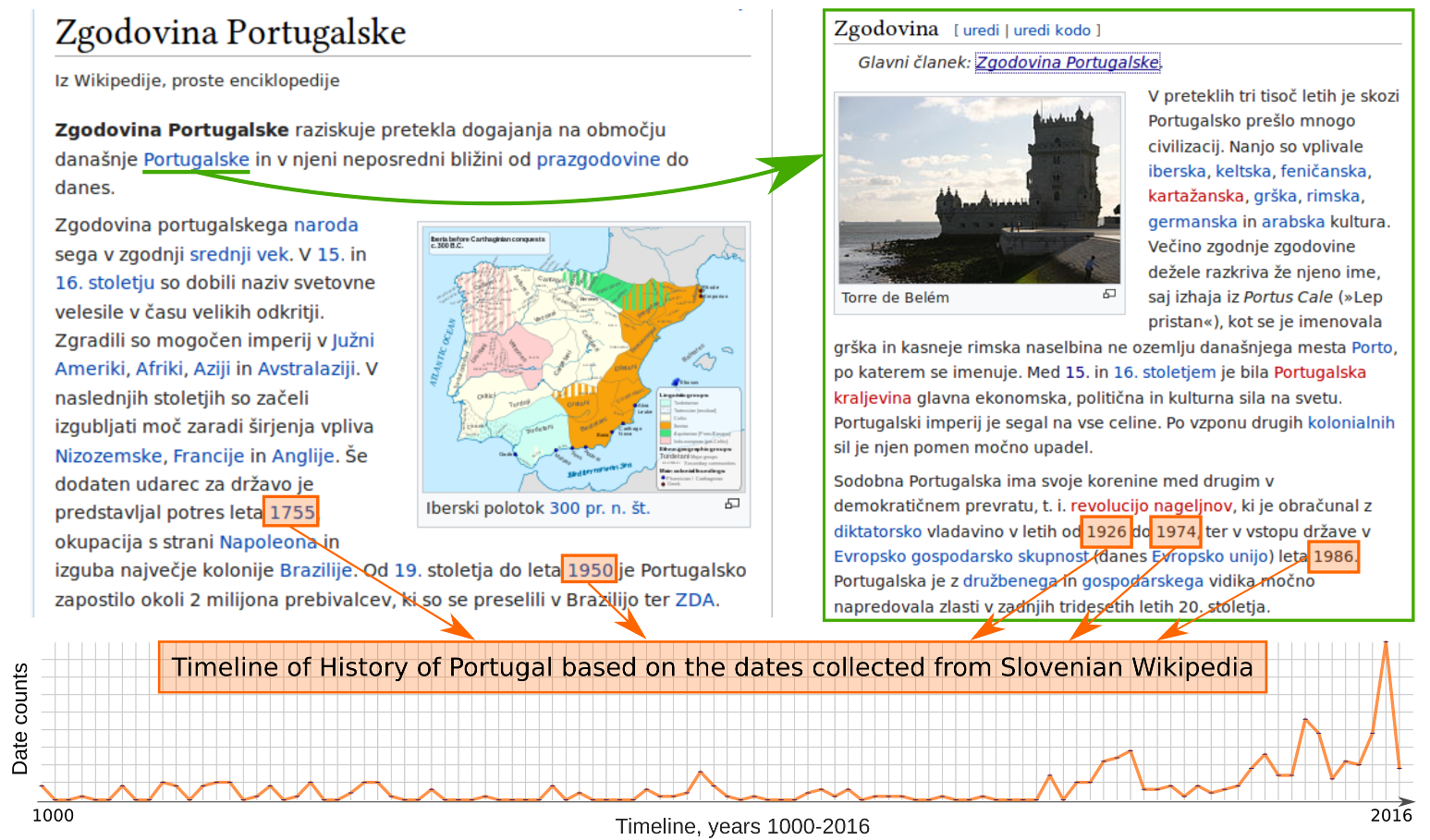}}
    \caption{\textbf{Data collection.} We show parts of the article on Portuguese history and one of its outlinks, as they appear in Slovenian Wikipedia in 2016. We collect all 4-digit numbers from the main text of the article and all its outlinks, and analyse the resulting distribution (bottom part of the figure). 
    }
    \label{fig:f1}
\end{figure}

\begin{figure*}[t!]
\centering
\subfloat[Extracted timelines: 30 language editions.]{
    \includegraphics[width=0.5\textwidth]{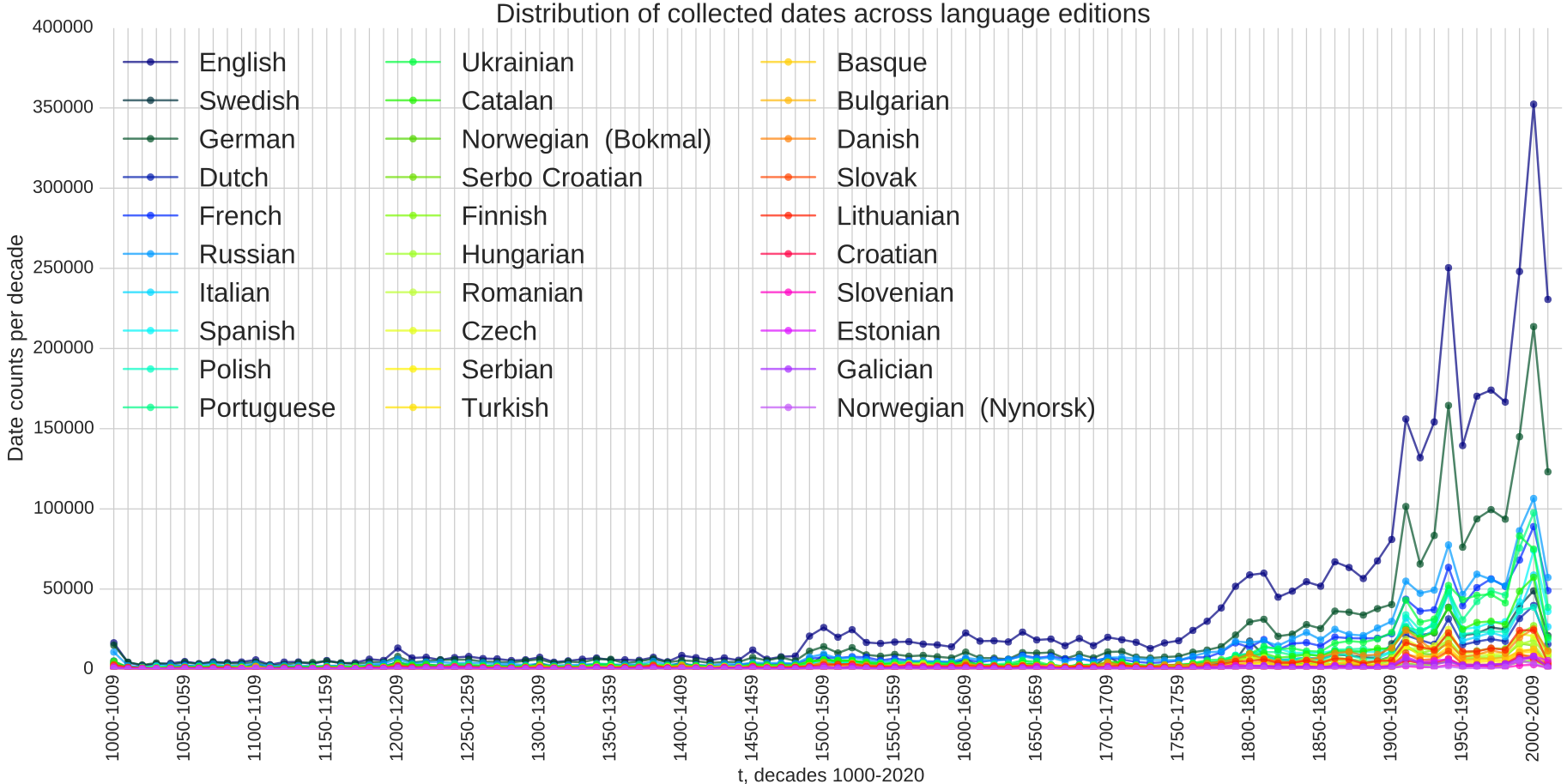}
    \label{fig:pool}
}
\subfloat [Extracted timelines: selected countries] {
    \includegraphics[width=0.5\textwidth]{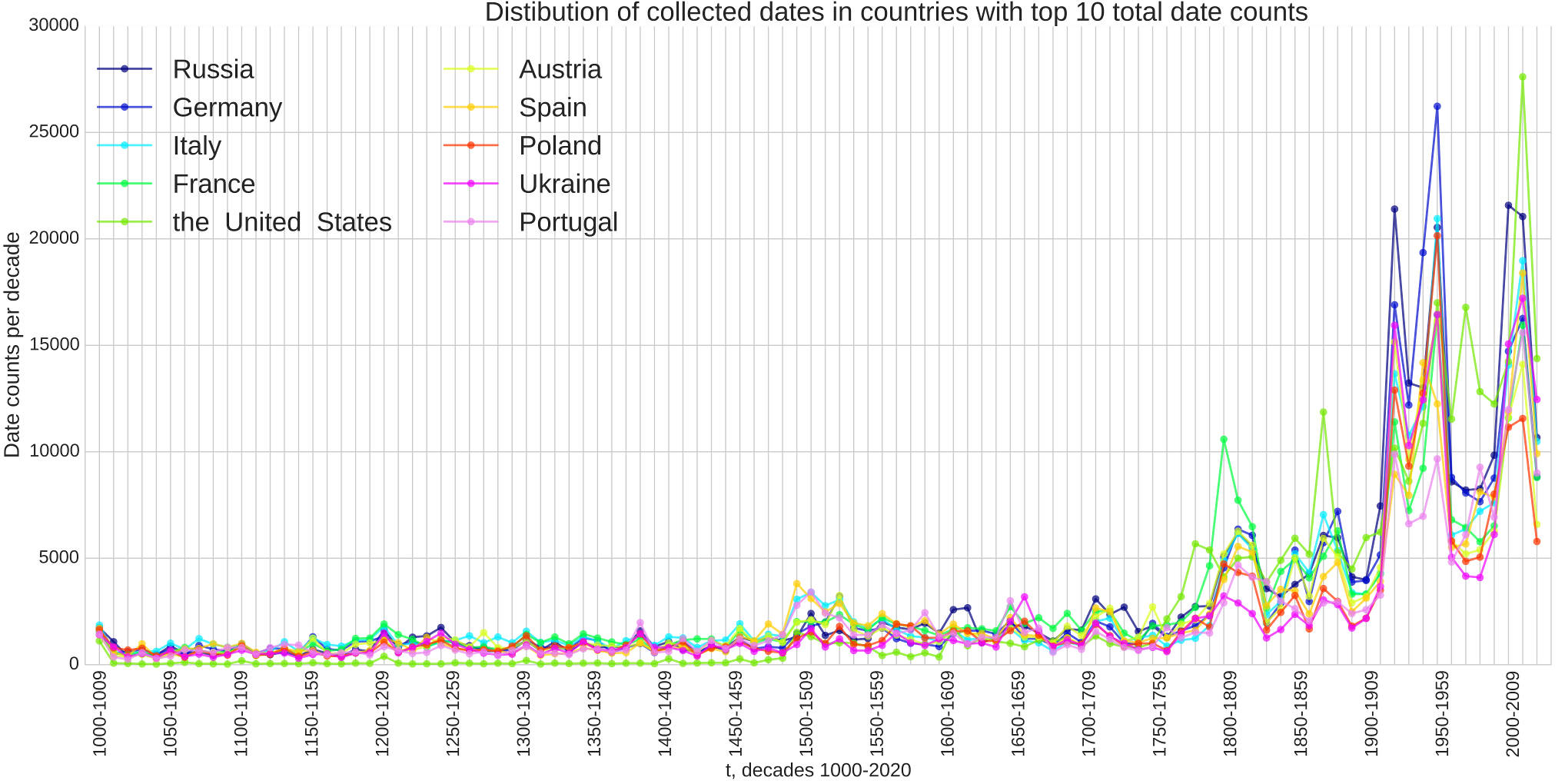}
    \label{fig:pool_countries}
}
\caption{\textbf{Distribution of collected dates.} \subref{fig:pool} in 30 language editions and \subref{fig:pool_countries} in top 10 countries according to the number of collected dates. Across editions of all sizes, and across countries we observe the same strong bias towards dates within the last 100 years, while dates from before 1500 are rarely mentioned.}
\label{fig1}    
\end{figure*}

\textbf{Data collection:} For each of the 193 countries we locate an article in the English edition of Wikipedia, titled 'History of X', where X is the country name. Using Wikipedia's inter-language links, we retrieve other language versions of the article from sister editions. We limit the analysis to 30 largest Wikipedia editions (more than 125,000 articles\footnote{Wikipedia:~List~of~Wikipedias,~ \url{http://en.wikipedia.org/wiki/List_of_Wikipedias} (accessed~Nov.~13,~2016)}) providing these languages are native to Europe. By applying this setup we avoid issues connected with extraction and alignment of dates from the languages using different calendars and alphabet systems. The limitations related to multilingual data retrieval and the choice of linguistic scope are discussed in detail in Section \ref{limitations}. 

We retrieve the main text of each article (as HTML, excluding text related to, e.g. references) from the English Wikipedia and -- if available -- from all 30 selected sister editions, as well as the text of all Wikipedia articles to which these pages link. 
We focus on the out-links because they are embedded in the main articles' texts and thus immediately available for a reader to inspect, unlike, for example, the in-links, which could not be found by reading the article page.
We find between 14,927 (Italy) and 394 (the Federated States of Micronesia) articles related to the history of each nation.
In order to assess the coverage of historical periods, we choose a language-independent measure -- the mentions of year numbers in the article text. Since we are interested in historical events of the last millennium, we retrieve all 4-digit numbers in the range between 1000 and 2016 from the main text of all articles in our collection. Fig. \ref{fig:f1} illustrates the process with an example of an article on the history of Portugal in Slovenian Wikipedia. In cases when paragraphs consist mostly of hyperlinks (more than 50\% of words are hyperlinks), we record no dates from them, since there is little narrative in such paragraphs.

We ran the data collection in July 2016, using the access provided by Wikimedia Tool Labs \footnote{Wikimedia~Tool~Labs,~\url{https://wikitech.wikimedia.org/}~(accessed~Nov.~13,~2016)} as well as the Wikipedia API \footnote{Wikipedia~API~for~Python,~\url{https://pypi.python.org/pypi/wikipedia/} (accessed~Nov.~13,~2016)}. We retrieved approximately 17M dates from 773,121 articles in 30 language editions. 

%%%%%%%%%%%%%%%%%%%%%%%%%   Data evaluation   %%%%%%%%%%%%%%%%%%%%%%%%%%%%%%%%%%%%%%%%
%%%%%%%%%%%%%%%%%%%%%%%%%%%%%%%%%%%%%%%%%%%%%%%%%%%%%%%%%%%%%%%%%%%%%%%%%%%%%%%%%%%%%%
\textbf{Data validation:}\label{eval}
In order to ensure internal reliability of our extraction method, we check whether the extracted numbers are years rather than numerals indicating, for example, height. We create a random sample of 3,300 extracted 4-digit numbers evenly split across 30 languages and 11 centuries, and ask 3 independent human coders to evaluate each case, i.e. to say whether a number is a date or not (false positive). For each language there are 110 evaluation tasks, which consist of: the potential date (4 digits), the text surrounding the potential date in the original language (40 characters before and after the number), and the same text translated into English via Google Translate (except for the English edition case). If the coder is unsure about a number, we treat it as a false-positive. Each case is settled by the majority vote. The inter-rater agreement is substantial (Fleiss' kappa = .77). We compute the expected error rates for centuries,
\begin{equation}\label{eq:E_c}
    <E_c> = \frac{1}{D_c} \sum_{l} (\frac{n_{c,l}} {10} d_{c,l}),
\end{equation}
and language editions,
\begin{equation}\label{eq:E_l}
    <E_l> = \frac{1}{D_l} \sum_{c} (\frac{n_{c,l}} {10} d_{c,l}),
\end{equation}
where $D_l$ and $D_c$ are the total count of collected potential dates per language and century, and $n_{c,l}$ is false positives count in our random sample of $d_{c,l}$ numbers collected per language edition $l$ and century $c$. 

We report the expected error rates for both centuries and language editions in Table \ref{table:ev}. All language editions and most of the centuries show a very low expected error rate (below .04). We estimate the highest error rate in the 11th century (.24), since a large number of extracted digits turned out to be numerals relating to heights, population counts, etc. This error is present mostly in this century, presumably due to the numeral 1000 being often used for other purposes than mentioning a year. Other false-positives include dates from Before Christ. In the more recent centuries our extraction method is very exact.

\begin{figure}[!ht]
    \centering
    \subfloat[Toy data]{ \label{fig1a}\includegraphics[height=.15\textheight]{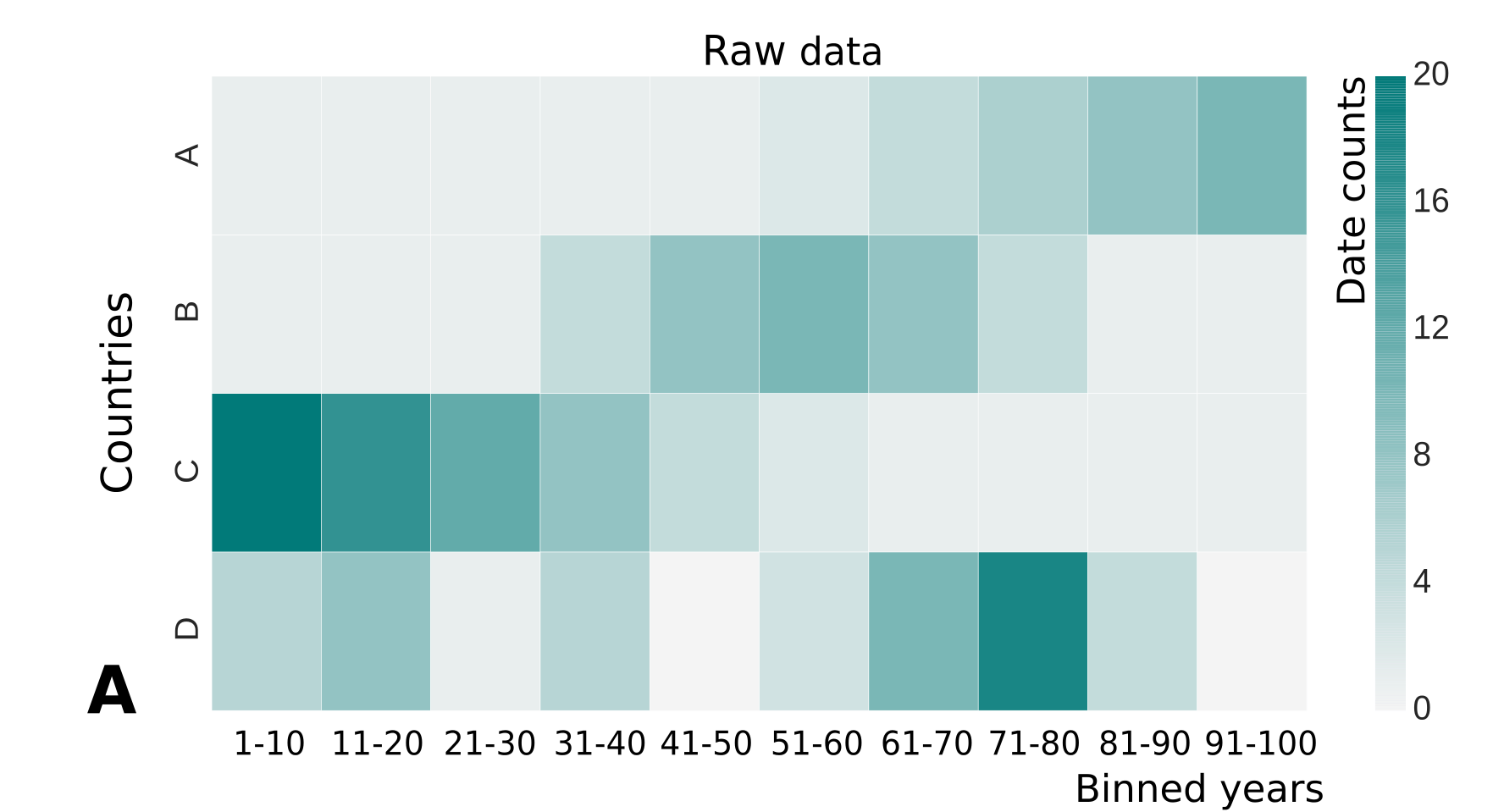}}
    \qquad
    
    \subfloat[Comparing toy data to the Null Model baseline]{\label{fig1b}\includegraphics[height=.15\textheight]{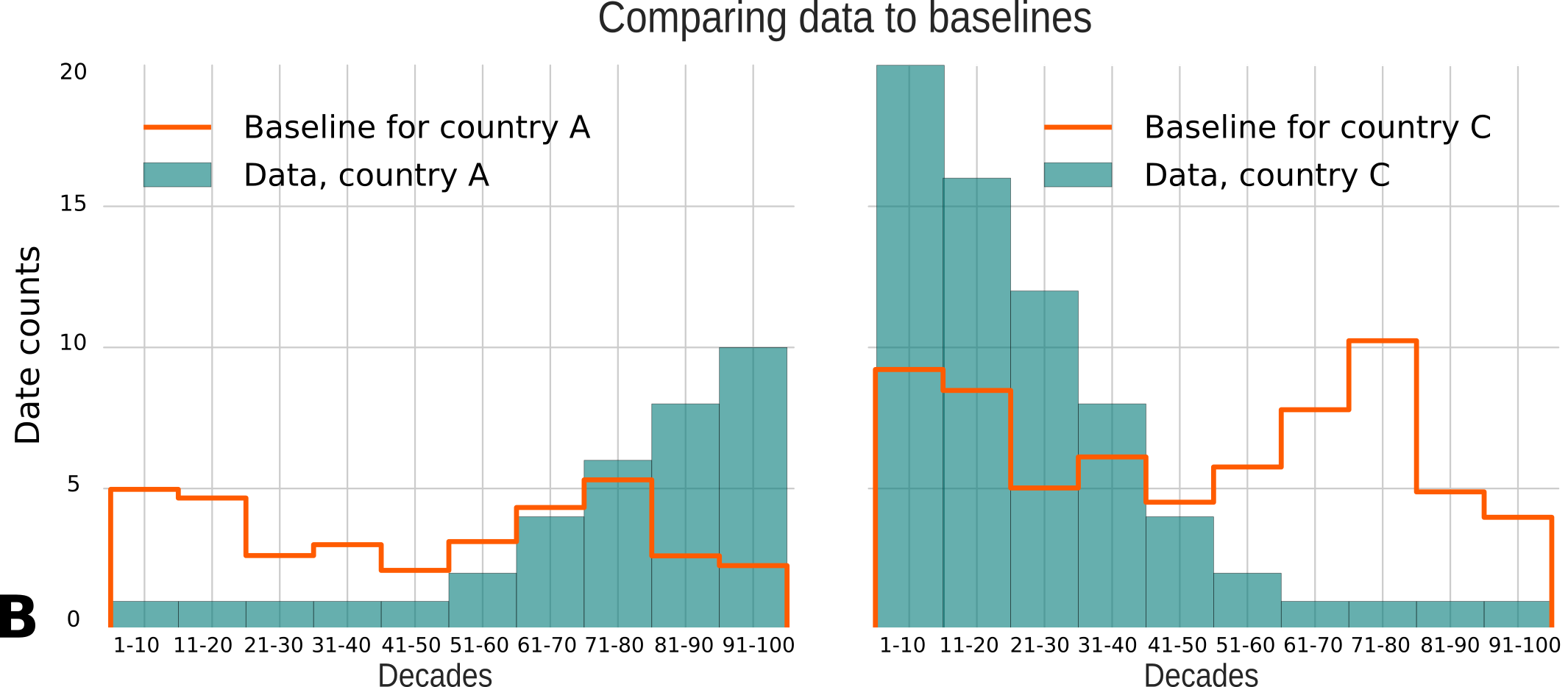}}
    \qquad
    
    \subfloat[Extracted focal points]{\label{fig1c}\includegraphics[height=.15\textheight]{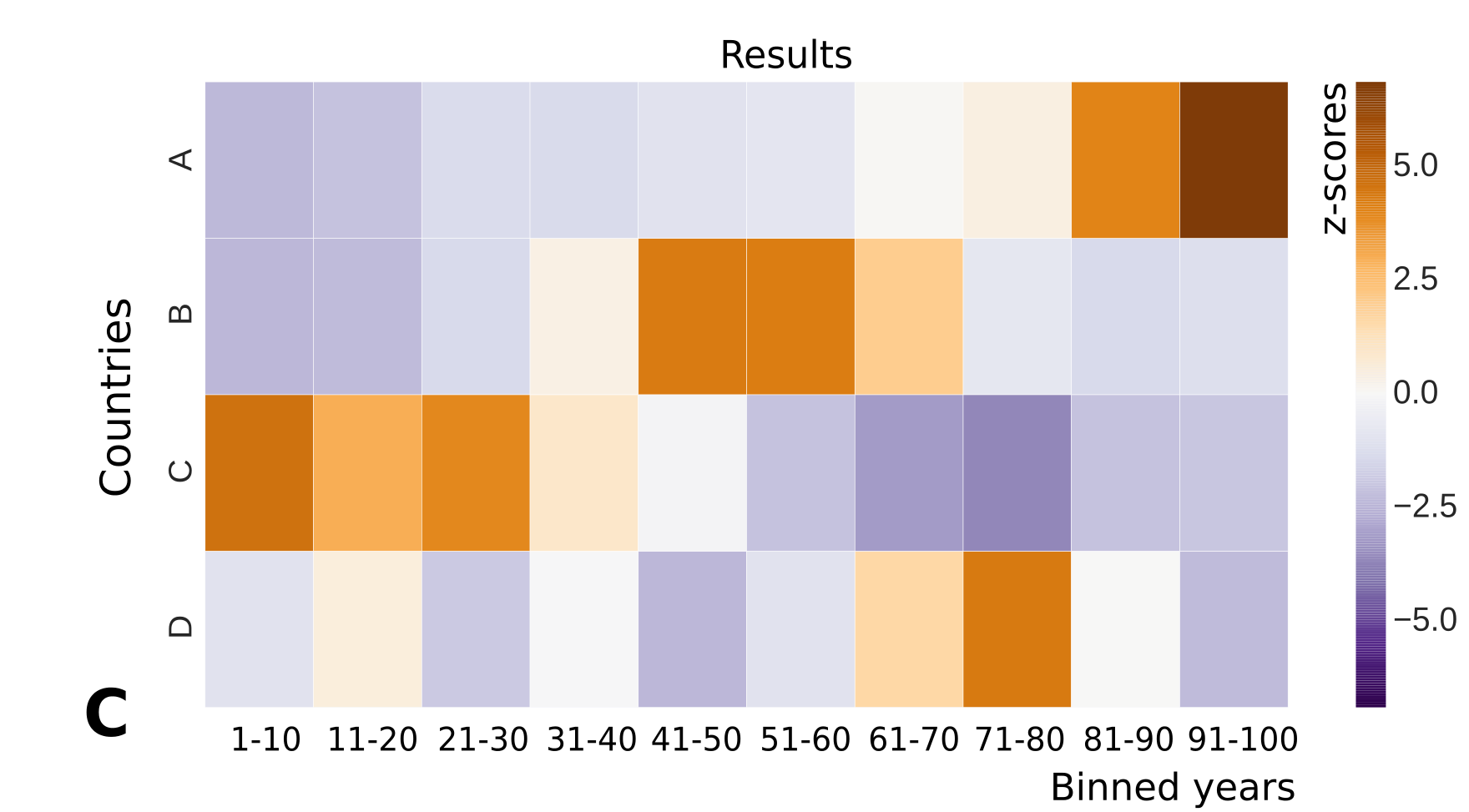}}

\caption{\textbf{Illustration of the method.} Fig.\ref{fig1a} shows the initial distribution of toy data for hypothetical countries A-D. The dates are binned into decades, each cell colored according to the number of dates. Fig. \ref{fig1b} illustrates the method on countries A and C. We plot a histogram of date counts for each country (green bins), and compare them with expected baselines (orange lines). The baselines are the average over four initial distributions, adjusted to match each country's total date count. Thus, the baselines are unique for every country and decade. Finally, in Fig. \ref{fig1c} we convert the differences between the data and the baseline into $z$-scores. }
\label{fig:fig1}
\end{figure}

\section{Approach and results}
%%%%%%%%%%%%%%%%%%%%%%%%%%%%%%%%%   RQ1   %%%%%%%%%%%%%%%%%%%%%%%%%%%%%%%%%%%%%%%%%%%%
%%%%%%%%%%%%%%%%%%%%%%%%%%%%%%%%%%%%%%%%%%%%%%%%%%%%%%%%%%%%%%%%%%%%%%%%%%%%%%%%%%%%%%
In this section, we describe our approach and present the findings regarding inter-lingual portrayal of national histories in Wikipedia.
The approach consists of three parts: \ref{sub1} identifying most covered historical periods across countries and languages, \ref{sub2} extracting the focal points of national histories across all language editions, and \ref{sub3} quantifying the amount of inter-language agreement on representation of national histories. 

\begin{figure*}[!ht]
\centering
    \includegraphics[width=\textwidth]{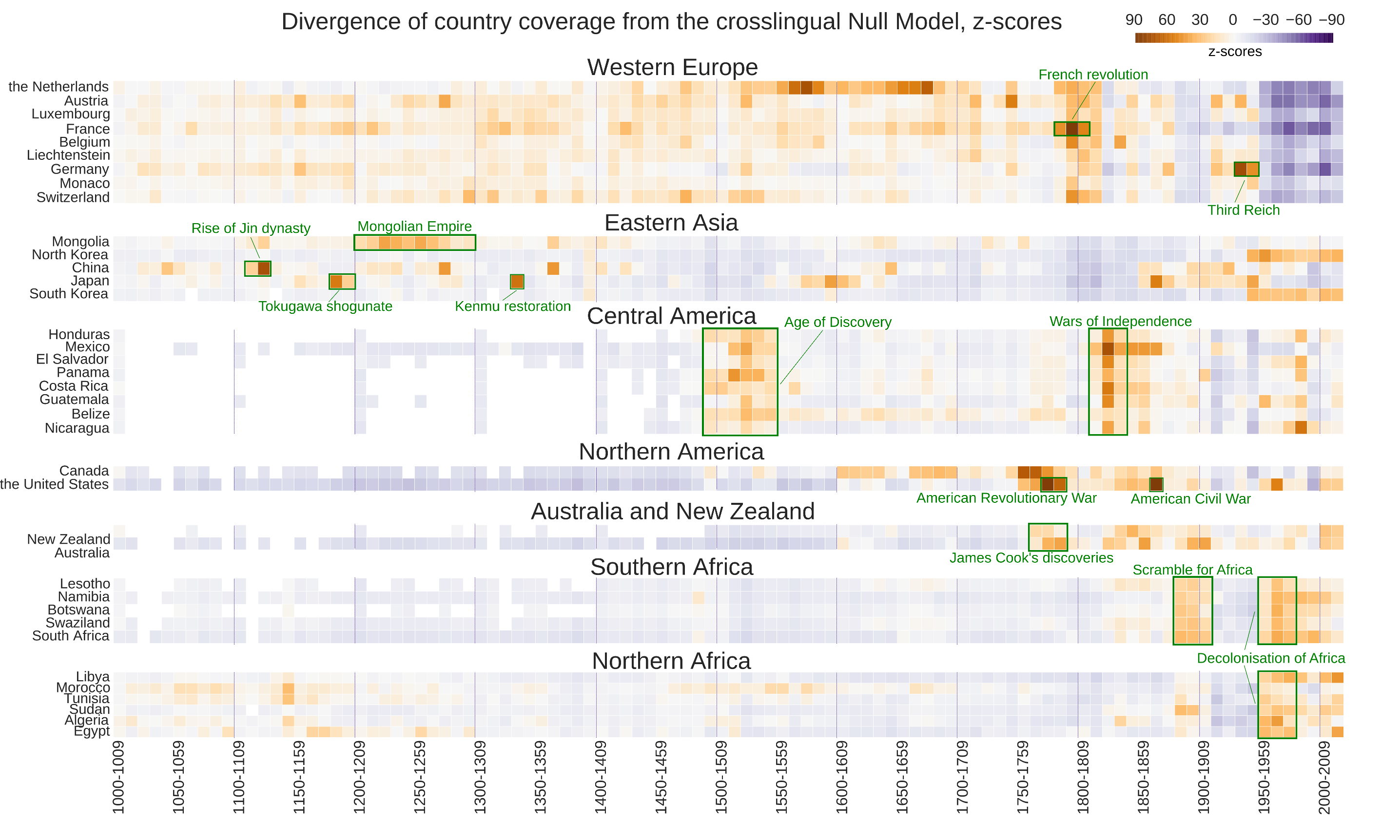}
    \captionof{figure}{\textbf{Temporal focal points of selected countries\protect\footnotemark.} 
    $z$-scores below -6 and above 6 correspond to Bonferroni-corrected $p$-values $<$ 0.01, which means the results in all coloured cells are statistically significant. Higher $z$-scores (orange) correspond to positive differences between the observed and the expected date count per decade. Cells with fewer than 30 dates are masked out. Interpretation of historical events corresponding to some focal points (in green) is offered by history experts. The distributions of focal points suggests there are similarities across countries within geopolitical blocs.
    \label{fig:z-continents}
    }
\end{figure*}

\subsection{Most covered historical periods}\label{sub1}
In order to gain a better understanding of our dataset, we first look into timelines of extracted dates. We present the distribution of collected dates across all 30 language editions (Fig. \ref{fig:pool}) and ten selected countries (Fig. \ref{fig:pool_countries})  -- those with the most available dates. 

\textbf{Results.} Across language editions, the data show a bias towards recent dates, having a large proportion of dates (between 60 and 80 percent) in the more recent decades (since 1800), and very low date counts before 1500. This is partly due to the chosen subject (nation states), but also points to a more general recency bias. Thus, Wikipedia readers can find a more detailed documentation of historic events of the past 200 years, compared to earlier centuries. Apart from intense coverage of the most recent events (2000s), we also observe peaks of date mentions that correspond to some of the most violent recent conflicts: Napoleonic war (1800-10s) and the First (1910s) and the Second (1940s) World Wars.

\subsection{Historiographic focal points of countries}\label{sub2}
We build a Null Model that describes expected frequencies of dates in each decade under the assumption that all countries are presented equally. By comparing the observed frequencies with the expected baseline, we are able to detect significant focal points. For this part of the analysis we aggregate all country-related dates across language editions. 

\subsubsection{Method: A Null Model of focal points.}
Our approach is essentially an urn model, with replacement and without duplication, based on the underlying Dirichlet-multinomial distribution. First, we create a pool $M$ of the dates found in all language editions for all countries. Then, for each country $i$, we randomly draw from the pool $N_i$ dates, where $N_i$ is the number of dates related to the history of country $i$ collected from all language editions. We then count how many of these dates fall in each decade. We repeat the process 1,000 times. For each decade we thus build a distribution of the number of dates it can contain, within the hypothesis of events randomly distributed in time. Furthermore, we can compare the mean of the expected distribution $E[w_{i}^{d}]$ with the empirical date count for the country in the same decade, $w_{i}^{d}$, and convert the difference into a $z$-score. This procedure allows us to identify for each country in which decades the number of observed dates $w_{i}^{d}$ differs significantly from the expected number of dates in this decade. Mathematically, the $z$-score of country $i$ in decade $d$ is given by:
\begin{equation}
  z_{i}^{d} = \frac{w_{i}^{d} - \mathrm E[w_{i}^{d}]}{\sigma_{i}^{d}},
  \label{eq:z_scores}
\end{equation}
where $\mathrm E[w_{i}^{d}]$ is the mean of the simulated date counts in decade $d$ across 1,000 random draws, and $\sigma_{i}^{d}$ is the standard deviation of the simulated date counts.

Fig. \ref{fig:fig1} illustrates the method with a toy example. We consider four hypothetical countries A-D with different artificial timelines (Fig. \ref{fig1a}). The dates are binned into decades, each matrix cell indicates the number of dates in the corresponding decade. Fig. \ref{fig1b} shows histograms of date counts per country. Orange lines correspond to expected distributions given by the Null Model, which are simply the average over four initial distributions, adjusted to match the total country date count. These baselines vary across decades and countries so that to account for differences in countries' total datecount. We then convert the differences between the observed data and the baselines into $z$-scores (Fig. \ref{fig1c}). For each country we can now extract the decades in which the number of collected dates differs significantly from the expected baseline. This method is especially useful when differences in counts are not directly comparable, as we have in our case: all row sums in Fig. \ref{fig1a} differ. To illustrate, in Fig. \ref{fig1a} the cell with the largest datecount is country C in the first decade, but after comparing with the expected baseline (Fig. \ref{fig1b}), country A in the last decade stands out most, although its underlying count is smaller.

\textbf{Analytical solution.}
As an alternative to the simulation approach that we use, the distribution of results of our Null Model could also be computed analytically for each decade and country. Drawing from an urn without replacement is described by a hypergeometric distribution, defined as:
\begin{equation}
    P(X=k_{c,d}) = H(M,K_d,N_i,k_d) = {{{K_d \choose k_d} {{M-K_d} \choose {N_i-k_d}}}\over {M \choose N_i}},
\end{equation}
where $M$ is the population size (total date count across countries and languages), $K_d$ is the number of successes (the date count in the decade across languages), $N_i$ is the number of draws (the date count for the country across languages), and $k_d$ is the number of positive outcomes (the date count in the decade predicted by our Null Model). $k_d$, the expected mean value of our Null Model, can be computed as $N_i \cdot \frac{K_d}{M}$, and the standard deviation is $\sqrt{N_i{K_d\over M}{(M-K_d)\over M}{M-N_i\over M-1}}$.
A $p$-value for the significance of a deviation given by this Null Model can be determined by using the cumulative distribution function of the hypergeometric distribution.

\begin{figure}
    \centerline{\includegraphics[height=.92\textheight]{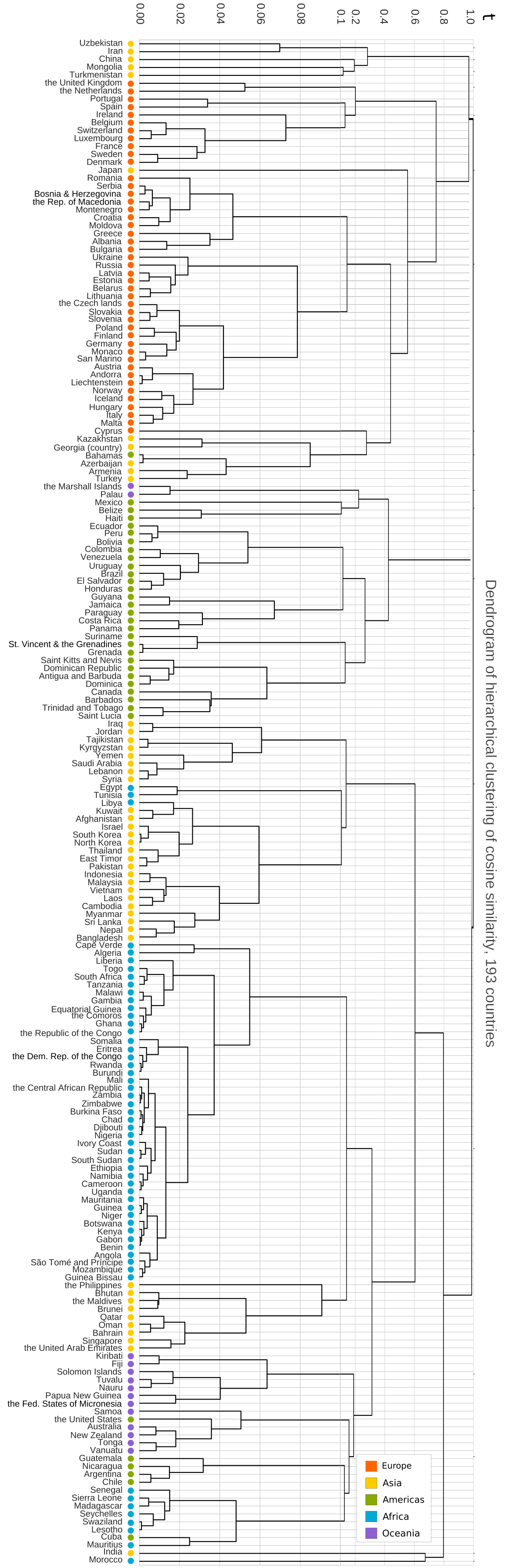}}
    \caption{\textbf{Complete dendrogram of hierarchical clustering}, based on cosine similarity values for all country pairs. The clusters of similarly largely correspond to geopolitical regions.}
    \label{fig:dendrogram}
\end{figure}

\textbf{Results.}
The results of this procedure for selected geopolitical blocs are reported in Fig.\ref{fig:z-continents}. $z$-scores below -6 and above 6 correspond to Bonferroni-corrected $p$-values $<$ 0.01 (the expected distributions of decade counts are approximately normal), which means the results those cells are statistically significant.
There are noticeable differences in distributions of focal points (in dark orange) across countries. For Western European countries, we observe high coverage of the Medieval and Early Modern periods (until $\sim$1800). Specific periods of interest for individual countries include, for example, the French Revolution in France (1780-90s) and the Third Reich in Germany (1930-40s).
By contrast, in East Asia the focal points are more heterogeneous. For Mongolia, the timeline focuses on the Mongolian Empire in the 13th century. Articles on Japanese and Chinese histories exhibit a strong focus on specific small time frames: the rise of the Tokugawa shogunate (1180-90s), the Kenmu Restoration (1330s) and the beginning of the Edo period (around 1600) in Japan; and the rise of the Jin (1120s), Yuan (1270s), Ming (1360s) and Qing (1640s) dynasties in China. Only with stronger European involvement in the region (starting in the mid-19th century) there is a more steady coverage.
For Central America, the timelines focus on the Age of Discovery (late 15th - early 16th centuries), and the Spanish-American Wars of Independence (first half of the 19th century).
In North America, the eras of the American Revolutionary War (end of 18th century) and the American Civil War (1860s) are most noticeable.
For different regions of Africa, historical timelines strongly focus on the periods of its occupation and colonisation (Scramble for Africa in late 19th century), and recent history following its decolonization in the 1960s.
In contrast to Southern Africa, North African national timelines focus on the Medieval history (Caliphate era), which is also the time of close interaction with Europe. The coverage seems to seize around 1300, just before the outbreak of the Black Death epidemic.
For Australia and New Zealand the peaks in 1760-80s correspond to the expeditions of James Cook discovering Oceania and South Pacific. Over the next centuries, as contacts between Europeans and the local population grew, the coverage remains stable.
 
Overall, the number of discovered `focal points' differs across regions. Within 30 examined Wikipedia editions, there is a disproportionate focus on histories of European countries, and the coverage of non-European states seems more intense in the periods when those states had closer interactions with Europe. 

\begin{figure}[t!]
    \centerline{\includegraphics[width=.5\textwidth]{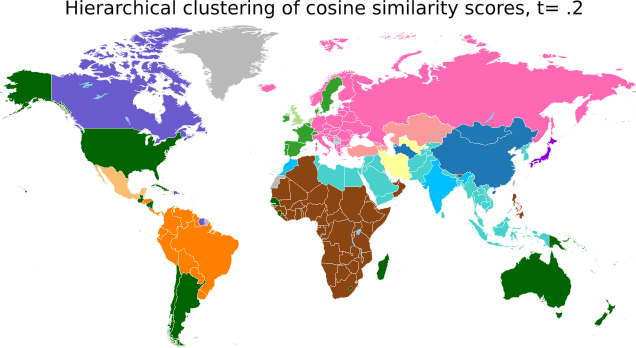}}
    \caption{\textbf{World map of country clusters\footnotemark[\value{footnote}].}~This results from cutting the dendrogram of hierarchical clustering at a threshold $t=.2$. The countries within coloured clusters have similar temporal focal points, based on the articles in analysed editions, and map well to geopolitical regions.}
    \label{fig:map}
    {\vskip-1ex}
\end{figure}

\textbf{Clustering.} We use the results reported in Fig.\ref{fig:z-continents} to group countries based on their historical timelines' similarity. We represent each country as a vector of $z$-scores, and group together the countries whose $z$-score values across decades are similar both in direction and intensity. We compute pairwise cosine similarity between all countries, and apply hierarchical clustering  with complete linkage \cite{mullner} on top of the obtained values. The resulting dendrogram (Fig. \ref{fig:dendrogram}) shows that the clusters correspond rather well to geopolitical regions. To illustrate this point, we cut the dendrogram at an (arbitrary) level $t$=.2, and plot the resulting 18 country clusters on a world map (Fig. \ref{fig:map}). It suggests that focal points of the countries from the same geopolitical regions are similar in the editions that we analyse. It is important to underline, however, that similar focal points do not necessarily indicate a reference to the same historical events.

\footnotetext{Additional plots illustrating the distribution of focal points across all countries, and the complete dendrogram of hierarchical clustering could be viewed at~\url{http://annsamoilenko.wixsite.com/homepage/historical-landscapes} }

Combining information about cluster membership in Fig.~\ref{fig:map} with the significant focal points presented in Fig. \ref{fig:z-continents}, we get a transnational impression on patterns of similarity among national timelines. We see, for example, that most of Africa maps to one cluster. Despite individual differences between country histories, in the analysed descriptions, history of the entire continent is reduced to the periods of its (de-)colonisation. Similarly, focal points of most of Central and South American countries are limited to the Age of Discovery and their Wars of Independence. On the contrary, Europe is separated into several clusters, as here the differences among the individual national timelines are more distinct. 
This gives an impression of how the entire world groups into regions based on the extracted focal points of individual countries. Also, it illustrates that for some parts of the world (e.g. Africa and parts of Americas), the analysed timelines show a reduced view of history.
%\vspace{-5mm}
\subsection{Quantifying inter-edition agreement}\label{sub3}
In this section we investigate if national historical timelines are consistent across languages. For that, we compute a measure of their divergence across Wikipedia editions.

\textbf{Method: Jensen-Shannon divergence.} 
Based on the extracted probability distributions of years, for each country we compute a matrix of pairwise inter-language dissimilarities, using the Jensen-Shannon (J-S) divergence \cite{JS-divergence}:
\vspace{-3mm}
\begin{equation}\label{JS_Divergence}
    % \centering
    \begin{multlined}
        J(p\parallel q) = \frac{1}{2} \Biggl[\sum_t p(t) \log_2 \Bigl(\frac{2p(t)}{p(t) + q(t)}\Bigr) + \\
        \sum_t q(t) \log_2 \Bigl(\frac{2q(t)}{p(t) + q(t)}\Bigr)\Biggr],\
    \end{multlined}
\end{equation}
where $p(t)$ and $q(t)$ refer to the probability of year $t$ in the language editions $p$ and $q$. The divergence $J(p\parallel q)\in[0,1]$, with 0 indicating complete overlap between the compared distributions. Differences across language-specific timelines of each country are summarised by a square $m\times m$ matrix, where $m$ is the number of extracted language editions (up to 30) covering the country's history. We summarise interlingual differences in two values: median and spread of the distribution of $J(p\parallel q)$ values per country, and present them in a scatterplot for ease of visual analysis.

\begin{figure}[t!]
    \centerline{\includegraphics[width=.5\textwidth]{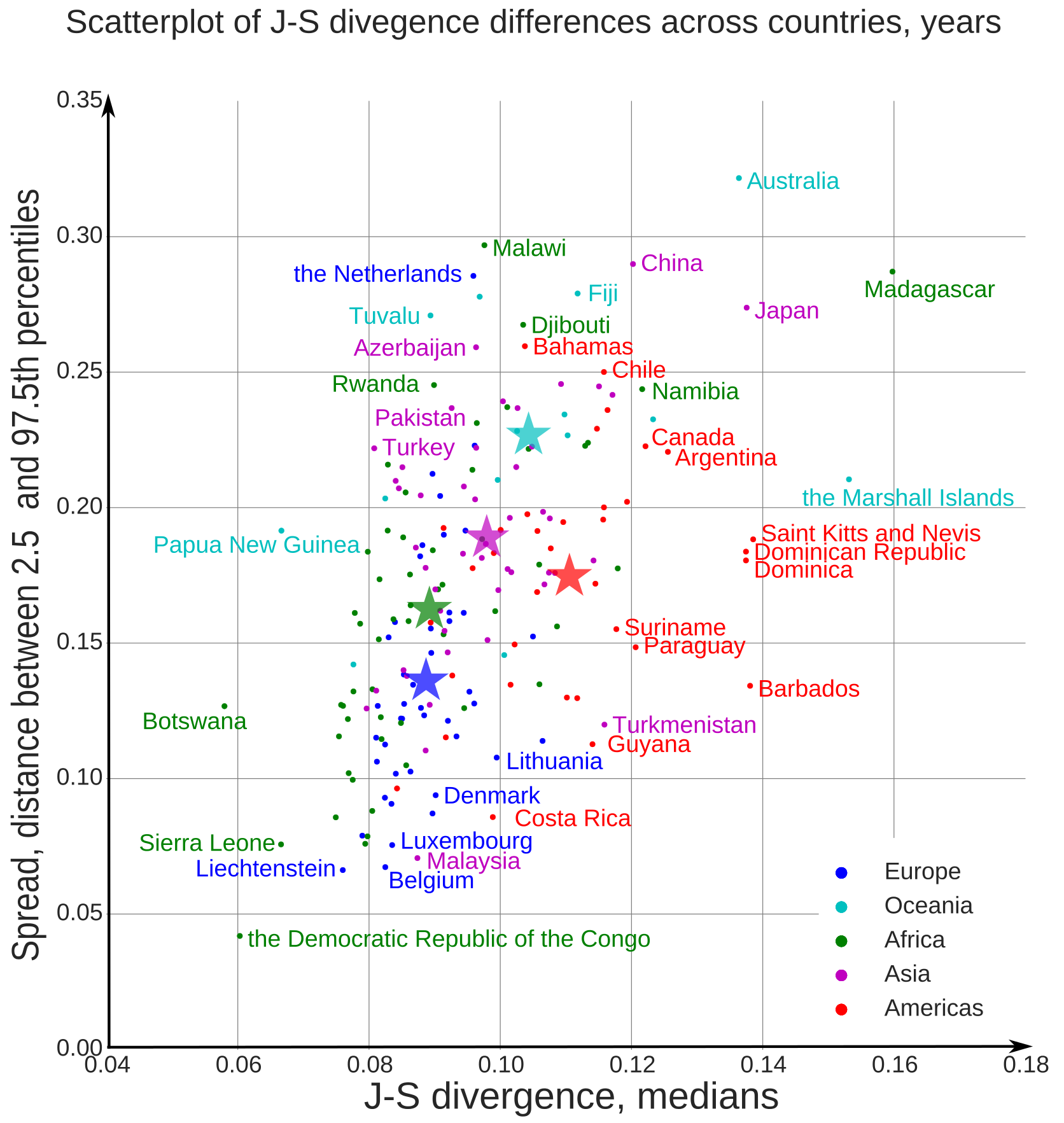}}
    \caption{\textbf{Inter-language consensus in Wikipedia articles on national histories,} based on pairwise Jensen-Shannon divergence values. Countries in the lower left of the plot show the highest consensus across editions.
    Stars represent data centroids for countries of the same region. The plot shows a high inter-language consensus on average, though the descriptions are not identical across editions. European countries exhibit the highest amount of consensus.}
    \label{fig:scatter}
\end{figure}

\begin{figure}[t!]
\centering
\subfloat[Belgium: high consensus]{
    \includegraphics[width=0.5\columnwidth]{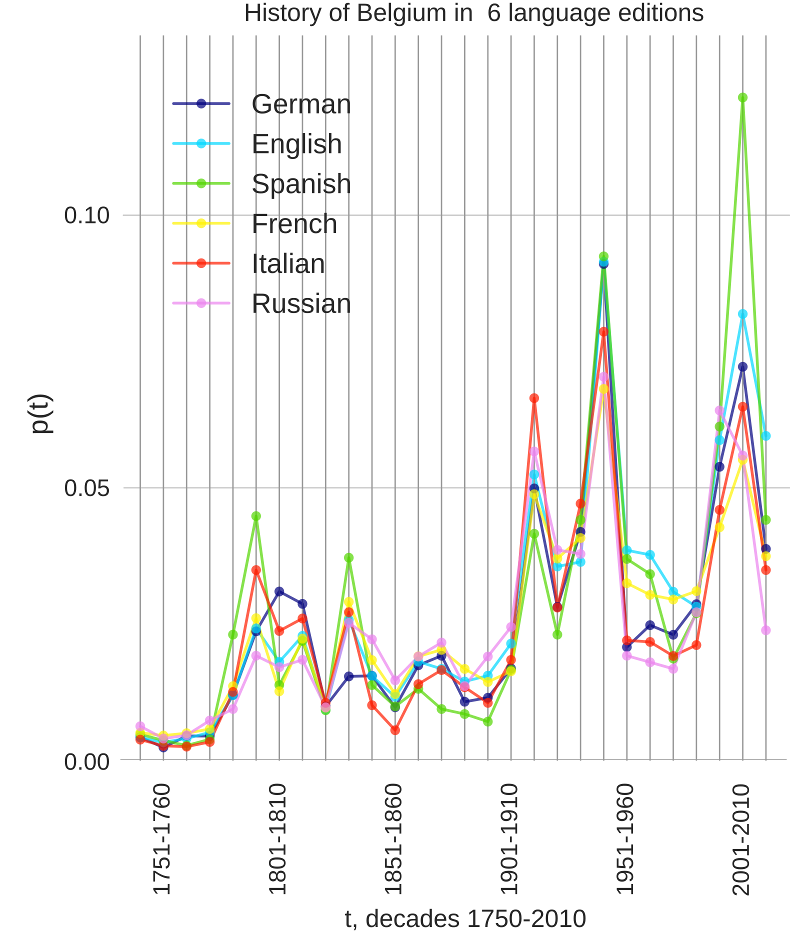}
    \label{fig:belgium}
}
\subfloat [China: low consensus] {
    \includegraphics[width=0.5\columnwidth]{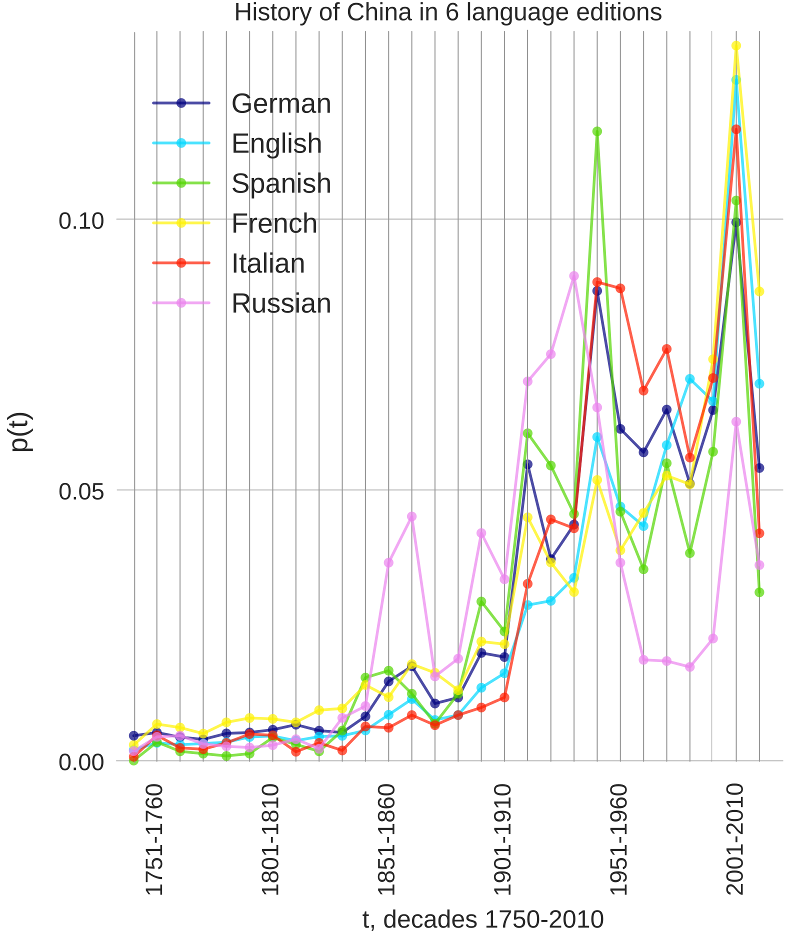}
    \label{fig:china}
}
\caption{\textbf{Inter-lingual consensus on histories of selected countries.} To illustrate cases with very high \subref{fig:belgium} and very low \subref{fig:china} inter-lingual consensus, we present parts of probability distributions of dates zoomed into 1750-2010s, for 6 large editions. Chinese timeline in Russian Wikipedia differs noticeably from the timelines in all other editions, while for Belgium all timelines are almost identical.}
\label{fig:examples}    
\end{figure}

\textbf{Results.}
The results of this approach are presented in Fig.\ref{fig:scatter}. 
Data points in the lower left quarter correspond to countries with the lowest medians and the narrowest distributions of J-S scores (i.e. the smallest differences between the most similar and the most different language pair), and thus, with the highest inter-lingual consensus. 
Overall, J-S scores are centered around very low values (medians between .06 and .16), which indicates a high average agreement across language editions. Their spread covers a higher range (up to .35), implying the presence of large differences between some language pairs. 
Based on the location of data centroids (stars), we observe higher interlingual consensus on the history of European and African countries, compared to Americas, Asia, and Oceania. The largest interlingual disagreement is found in the articles on the history of Australia, Malawi, Madagascar, China, Japan, and the Netherlands; some with the highest consensus are Liechtenstein, Belgium, the Democratic Republic of Congo, and Malaysia. 
In case of China (Fig. \ref{fig:china}), for example, high disagreement is partially driven by the differences between Russian and other language editions. This is especially evident during the period of Sino-Soviet split in 1960-80s, which is less densely covered in the Russian language Wikipedia. Timelines of history of Belgium, on the other hand, are almost identical across all 30 language editions (in Fig. \ref{fig:belgium} we present only 6 largest editions in order not to obstruct the view). Overall, we conclude that country-specific historical timelines differ across language editions, although on average such differences are rather small. 

%%%%%%%%%%%%%%%%%%%%%%%%%%%%%  Discussion  %%%%%%%%%%%%%%%%%%%%%%%%%%%%%%%%%%%%%%%%%%%
%%%%%%%%%%%%%%%%%%%%%%%%%%%%%%%%%%%%%%%%%%%%%%%%%%%%%%%%%%%%%%%%%%%%%%%%%%%%%%%%%%%%%%
\section{Discussion}
In this section we reflect on characteristics of the proposed computational approach, and discuss our empirical findings.

\textbf{Approach:} 
We have selected year mentions as a language-independent unit of analysis that can act as a measure of emphasis on certain historical events (this formalisation has been used before by Michel et al. \shortcite{Michel}). In principle, it is possible to choose any other quantifiable unit, such as mentions of geographical locations, persons, or events, providing there is an extra step for tackling inter-lingual entity disambiguation. Although the building blocks of our approach are not new in computational fields, they are new to the community of quantitative historians, where hitherto unseen inflow of newly digitised historical records pose a methodological challenge. Our approach is general enough to be applied to many large digitised datasets (such as demographic and economic records, census data, books, etc.), and it is suitable for comparative analysis of any number of countries across languages. By applying purely computational and data-driven methods, we are able to eliminate the bias that could be posed by the researcher's cultural background \cite{ailon}, and perform transnational analysis on a scale previously unknown to comparative historiography.

\textbf{Empirical results:} 
We apply our approach to a case study and investigate, what readers of different languages can learn about national histories from 30 Wikipedia editions. Our empirical results indicate the presence of \textit{recency bias} across language editions and countries: most retrieved dates belong to the recent decades, while those before 1500 are very sparse. Other studies report similar findings: a survey of students from Europe, the US, and Japan about the events that they perceived most important in the last 1,000 years showed that 60\% of the mentioned events happened in the last 300 years \cite{rovira}, while in our case it is between 60 and 80\% depending on the language edition. Recency bias is a well-known concept in the fields of social/collective memory and psychology of history, and it is sometimes referred to as genealogical \cite{candau}, autobiographical \cite{wertsch}, or, most commonly, communicative memory \cite{Assmann-2,Assmann-1}. We find that recent memories are actively documented on Wikipedia; possibly, because we simply know more about the recent past. However, like \cite{Pentzold-12}, we also find evidence that these narratives stretch beyond the limited domain of communicative memory (`floating gap' of 3-4 generations), and reflect long-term, stabilised cultural memory \cite{Assmann-2}. While recency bias is common in oral accounts of history, it is novel to demonstrate it for the context of a written encyclopedia.

The analysis of historiographic focal points of countries indicates inhomogeneous coverage across the continents, but high similarity within geopolitical country blocs. We find a multitude of focal points distributed across the whole timelines of European countries, while we observe very sparse coverage and no focal points in pre-Columbian Americas and Oceania. Significant focal points in non-European states appear to relate to the periods which are culturally and historically important for Europe, such as the discovery of Latin America and the Polynesian islands by European travellers, the beginning of European trade with China, and the period of close interaction between Europe and Northern Africa up until the Black Death epidemic. We interpret this as evidence of \textit{Eurocentric bias}, an issue well-documented in professional historiography \cite{geyer1995}, but also present in public perceptions of history, as cross-cultural surveys show \cite{rovira,liu2005}. Thus we find that Wikipedia, despite offering a democratised way of writing about history, reiterates similar biases that are found in the `ivory tower' of academic historiography.  Given that the focus of our analysis is on languages spoken in Europe, some dominance of Eurocentric perspectives is expected. Still, some languages that we study (e.g., English and Spanish) are widely spoken in other regions, such as Latin America and Africa.
Considering their international reach and the collaborative, global nature of Wikipedia, it is surprising to empirically confirm this imbalance towards European countries.

We also find \textit{high consensus across the examined editions} in describing individual country histories. Across language editions, extracted dates also peak in the same decades, which correspond to periods of highly violent conflicts. Although this finding is not immediately intuitive, previous studies have reported high consensus in how different cultures view important historical events \cite{rovira}. The authors explained this by the possible existence of cross-cultural collective memory, dominant hegemonic beliefs about the world history, and the narrowing cultural and interest differences between the communities. 
The latter has also been studied in Wikipedia context, finding that both linguistic \cite{Samoilenko} and geo\-graphic communities \cite{karimi} of Wikipedia editors are interested in similar article topics. We add to this research by demonstrating that in the case of history, the content (recovered timelines) of articles is also very similar across languages. 
These similarities in the public perceptions of history might be a result of a converging approach to history education, common exposure to media and entertainment (such as popular history TV shows), or lack of exposure to alternative historical material \cite{conrad}. Additionally, cross-lingual Wikipedia editors and bots might be responsible for inserting similar material in different language versions of the article \cite{hale}, which might be a factor in the similarities we find.

%%%%%%%%%%%%%%%%%%%%%%%%%%%%%  Limitations  %%%%%%%%%%%%%%%%%%%%%%%%%%%%%%%%%%%%%%%%%%
%%%%%%%%%%%%%%%%%%%%%%%%%%%%%%%%%%%%%%%%%%%%%%%%%%%%%%%%%%%%%%%%%%%%%%%%%%%%%%%%%%%%%%
\section{Limitations}\label{limitations}
Our methodological choices have implications on the conclusions we can draw from the analysis. Below we present the limitations of this study grouped by type.

\textbf{Unit of analysis:} One of the main limitations of any computational approach is the fact that it is reductional:  
in this study, we reduce the complexity of historiography to year mentions. The advantage of this approach is in focusing on an objective, quantifiable unit that can be compared across language editions without the biases introduced by translation. We leave other, language-specific methods of analysing historical narratives and sentiments outside of the scope of this study. Our measure, a year, is a very fine-grained unit when examining a millennium of human history. Still, by leaving mentions of decades or centuries unaccounted for, we loose some precision, especially for certain countries and earlier time periods where the exact dates of events are not well-known or documented. 

\textbf{Focal points:} We define focal points as time periods of significantly high mentions, compared to a random expectation model. Other formulations of Null Models are possible, which could describe a random process otherwise, and potentially result in non-identical outcomes. However, our model makes the least assumptions about the countries' histories and thus makes a good baseline. Interpretations of historical events related to some extracted focal points depict a viewpoint of selected history experts and are subjective.

\textbf{Linguistic scope:} We focus on 30 largest editions of Wikipedia which are also the languages native to the geographic region of Europe. For these editions, year is an acceptably robust unit of analysis, since these languages generally share date and time notation standards. We exclude other large editions such as Chinese, Arabic, and Farsi since their distinctive calendar- and numeral systems require developing language-specific methods of dates extraction, and this task goes beyond the scope of this study. The conclusions of our study should not be generalised to the whole Wikipedia, and are only valid for the studied editions.

\textbf{Multilingual data retrieval:} Our method of retrieving sister-articles from non-English language editions relies on Wikipedia's inter-language links (ILLs). Although the quality of ILLs is a debatable issue, studies have shown that the proportion of bidirectional ILLs between English and the largest European languages is around 98\% \cite{missing_ills}. We do not exclude the possibility that some of the multilingual articles might have been missed, however it is reasonable to assume that their absolute share will not have dramatically affected the results of the study.

\textbf{Article disambiguation:} Our analysis focuses on the articles with the specific title wording, 'History of X'. To solve title disambiguation issues in the English edition, we manually map all countries to corresponding Wikipedia articles on their modern history. In cases when a territory has changed names several times (having been a part of several countries, e.g., post-Soviet bloc), there might be multiple Wikipedia articles related to its history. We partially tackle this issue by including out-linked articles in our dataset, which in itself may introduce additional noise, and may potentially impact the findings.

\textbf{Inhomogeneous data:} We compare language editions at different ages, states of saturation, and sizes of underlying potential editor populations. This unavoidably leads to an overrepresentation of larger editions, for example, the pool of dates in RQ2 (Section \ref{sub2}) is heavily influenced by German and English editions.

\textbf{Data validity:} Data validation has shown high accuracy of our date extraction method. This is possible because we limited our analysis to the articles evidently related to history. The precision of the method might suffer when analysing texts of broader scope or focusing on the dates from Before Christ era. Already in our sample, we find small numbers of false-positives, e.g. 4-digit numerals expressing heights, lengths, or population counts. Although suitable for the current setup, our dates extraction method might need improvement if applied to a different dataset.

\section{Conclusions}
Our study explores Wikipedia as a novel data source for the community of historiographers. We focus on multilingual Wikipedia articles about country histories, and study popular perceptions of the past which are collaboratively created by non-professional history enthusiasts.

For this study, we have developed a computational approach which is scalable, language-independent, and flexible in terms of choosing its unit of analysis. We have used the approach to extract and visualise the periods of significant historical importance of 193 countries over the last 1,000 years, based on Wikipedia articles about their history in 30 language editions. We find that public narratives about history on Wikipedia are skewed towards more recent events, and are distributed unevenly across the continents, with significant focus on the history of European countries. We also find that cross-lingual consensus on national historical timelines is on average rather high, although disagreements are also present. 

The observed `peaks' and `lows' of interest to certain time periods, as well as cross-lingual differences in national timelines, might have different explanations. If these dissimilarities are \textit{intentional}, they might be a reflection of cultural differences, and in this case, our results could be interesting for historians and cultural scientists who might wish to explore the topic in greater detail and with other methods. If these differences are \textit{accidental} or could be reduced to `missing data', our findings could be actionable for the Wikimedia community and enthusiastic editors who wish to improve the quality of the articles in various language editions. In any case, our results show that Wikipedia's historical reference articles are not free from gaps and biases. We hope that History teachers and students, as well as lay readers who use Wikipedia to enrich their knowledge about world history, would benefit from this awareness.

\section{Acknowledgements}
The authors would like to express their sincere gratitude to the members of GESIS CSS Data Science team, and in particular, to Mohsen Jadidi, Fariba Karimi, Mathieu G{\'e}nois, and Sebastian Stier for insightful discussions of the project during its earlier stages.

\bibliographystyle{aaai}
\fontsize{9.0pt}{10.0pt}
\bibliography{sample}

\end{document}